\documentclass[prl,twocolumn,letterpaper,showpacs,groupedaddress,superscriptaddress,nofootinbib,floatfix,preprintnumbers,tightenlines]{revtex4}
\usepackage{hyperref,amssymb,amsmath,graphicx,xcolor,cancel}
\usepackage{bm}

\def\si{^1 \hskip -0.03in S _0}
\def\siii{^3 \hskip -0.025in S _1}

\def\sigmaPHYS{332.4({\tiny \begin{array}{l}
		+5.4 \\ - 4.7
	\end{array}})}
\def\sigmaEXPT{334.2(0.5)}

\begin{document}


\title{{ Ab initio} calculation of the  $np\rightarrow d\gamma$ radiative capture process}

\author{Silas~R.~Beane} 
\affiliation{Department of Physics,
  University of Washington, Box 351560, Seattle, WA 98195, USA}
  
\author{Emmanuel~Chang}
\affiliation{Institute for Nuclear Theory, University of Washington, Seattle, WA 98195-1560, USA}

 \author{William Detmold} \affiliation{
Center for Theoretical Physics, 
Massachusetts Institute of Technology, 
Cambridge, MA 02139, USA}
 
\author{Kostas~Orginos}
\affiliation{Department of Physics, College of William and Mary, Williamsburg,
  VA 23187-8795, USA}
\affiliation{Jefferson Laboratory, 12000 Jefferson Avenue, 
Newport News, VA 23606, USA}

\author{Assumpta~Parre\~no}
\affiliation{Dept. d'Estructura i Constituents de la Mat\`eria. 
Institut de Ci\`encies del Cosmos (ICC),
Universitat de Barcelona, Mart\'{\i} Franqu\`es 1, E08028-Spain}

\author{Martin J. Savage}
\affiliation{Institute for Nuclear Theory, University of Washington, Seattle, WA 98195-1560, USA}

 \author{Brian C. Tiburzi} 
\affiliation{ Department of Physics, The City College of New York, New York, NY 10031, USA }
\affiliation{Graduate School and University Center, The City University of New York, New York, NY 10016, USA }
\affiliation{RIKEN BNL Research Center, Brookhaven National Laboratory, Upton, NY 11973, USA }

\collaboration{NPLQCD Collaboration}

\date{\today}

\preprint{INT-PUB-15-010}
\preprint{NT@UW-15-02}
\preprint{MIT-CTP-4666}


\begin{abstract}
Lattice QCD calculations of two-nucleon systems are used to isolate the
short-distance two-body electromagnetic contributions to the radiative capture process $np\rightarrow d\gamma$,
and the photo-disintegration processes $\gamma^{(*)}d\rightarrow np$. 
In nuclear potential models, such contributions are described by 
phenomenological meson-exchange currents, while in the present work, 
they are determined directly from the quark and gluon interactions of QCD.
Calculations of neutron-proton energy levels in 
multiple background magnetic fields are performed at two values of the 
quark masses, corresponding to pion masses of $m_\pi\sim 450$ and 806 MeV, and are combined with pionless nuclear effective field theory to
 determine these low-energy inelastic processes. Extrapolating to the physical pion mass,
a cross section of $\sigma^{\rm lqcd} (np\rightarrow d\gamma)=\sigmaPHYS~{\rm mb}$ is obtained 
at an incident neutron speed of $v=2,200~{\rm m/s}$, consistent with the experimental value of
$\sigma^{\rm expt}(np\rightarrow d\gamma)=\sigmaEXPT~{\rm mb}$.
\end{abstract}

\maketitle

The radiative capture process, $np\rightarrow d\gamma$, plays a critical 
role in big-bang nucleosynthesis (BBN) as it is the starting point for the 
chain of reactions that form most of the light nuclei in the cosmos. Studies of radiative
capture \cite{Cox1965497,Tomyo2003401,Nagai:1997zz}, and the inverse processes of 
deuteron electro- and photo-disintegration, $\gamma^{(*)}d\rightarrow np$ 
\cite{Hara:2003gw,Moreh:1989zz,Schreiber:2000tb,Tornow:2003ze}, have 
constrained these cross-sections and have also provided critical insights into the interactions 
between nucleons and photons. They conclusively show the importance of 
non-nucleonic degrees of freedom in nuclei, which arise from meson-exchange currents 
(MECs) in the context of nuclear potential models \cite{Riska:1972zz,Hockert:1974qt}.
Nevertheless, in the energy range relevant for BBN, experimental investigations are challenging \cite{Ando:2005cz}.
For the analogous weak interactions of multi-nucleon systems, considerably less is known
from experiment but these processes are equally important. The weak two-nucleon 
interactions currently contribute the largest uncertainty in calculations of the rate for proton-proton 
fusion  in the Sun~\cite{Kong:2000px,Butler:2001jj,Ando:2007fh,Adelberger:2010qa,Chen:2012hm,Marcucci:2013tda,Rupak:2014xza}, and in neutrino-disintegration of the 
deuteron~\cite{Butler:2000zp}, which is a critical process needed to
disentangle solar neutrino oscillations. 
Given the phenomenological importance of electroweak interactions in light nuclei, 
direct determinations from the underlying theory of strong interaction, quantum chromodynamics (QCD),
are fundamental to future theoretical progress. Such determinations are also of significant phenomenological importance for calibrating long-baseline neutrino experiments 
and for investigations of double beta decay in nuclei.
In this Letter, we take the initial steps towards meeting this challenge and present the first lattice QCD (LQCD) calculations of the $np\to d\gamma$ process.
The results are in good agreement with experiment and show that QCD calculations
of the less well-determined electroweak processes involving light nuclei are within reach.
Similarly, the present calculations open the way for QCD studies of light nuclear matrix elements of scalar \cite{Beane:2013kca} (and other) currents relevant for dark matter direct detection experiments and other searches for physics beyond the Standard Model.

The low-energy cross section for  $np\rightarrow d\gamma$ 
is conveniently written as a multipole expansion in the 
electromagnetic (EM) field~\cite{Bethe:1950jm,Noyes:1964fx},
\begin{eqnarray}
\sigma(np\rightarrow d\gamma) & = & 
{e^2(\gamma_0^2+|{\bf p}|^2)^3\over M^4 \gamma_0^3 |{\bf p}|}
|\tilde X_{M1}|^2  + ...
\ ,
\label{eq:npdgsigma}
\end{eqnarray}
where 
$\tilde X_{M1}$ is the $M1$ amplitude,
$\gamma_0$ is the binding momentum of the deuteron,
$M$ is the mass of the nucleon, 
and ${\bf p}$ is the momentum of each incoming nucleon in the center-of-mass frame. 
The ellipsis denotes the contribution from $E1$ and higher-order multipoles 
(higher multipoles can be included systematically and improve the reliability of the description~\cite{Rupak:1999rk}, but are not
relevant at the level of precision of the present work).
In a pionless effective field theory expansion~\cite{Kaplan:1998sz,Kaplan:1998we,vanKolck:1998bw}, 
employing dibaryon fields to resum 
effective range contributions~\cite{Kaplan:1996nv,Beane:2000fi},
the leading-order (LO) and next-to-leading order (NLO)
contributions lead  to the $M1$ amplitude~\cite{Beane:2000fi,Detmold:2004qn}
\begin{eqnarray}
\tilde X_{M1} & = & 
{Z_d\over  -{1\over a_1} + {1\over 2} r_1 |{\bf p}|^2 - i |{\bf p}|}
\label{eq:npdgMone}
\\
&& \times \left[\ {\kappa_1 \gamma_0^2\over \gamma_0^2 +|{\bf p}|^2}\left( \gamma_0 - {1\over a_1} + {1\over 2} r_1 |{\bf p}|^2 \right)
+ {\gamma_0^2\over 2} l_1
\right] ,
\nonumber
\end{eqnarray}
where 
$\kappa_1=\left(\kappa_p-\kappa_n\right)/2$ is the isovector nucleon magnetic moment,
$Z_d = 1/\sqrt{1-\gamma_0 r_3}$ is the square-root of the residue of the deuteron propagator at the pole 
with $r_3$ the effective range in the $\siii$ channel, and 
$a_1, r_1$ are the scattering length and effective range in the $\si$ channel.
The quantity $l_1 = \tilde l_1 - \sqrt{r_1 r_3} \kappa_1$ encapsulates the 
short-distance two-nucleon interactions through $\tilde l_1$, but also depends on $\kappa_1$.
It is well established that gauge-invariant EM two-nucleon 
interactions (and direct photon-pion couplings in pionful effective field theories)~\cite{Kaplan:1998sz,Kaplan:1998we,Chen:1999tn,Chen:1999vd,
	Butler:1999sv,Rupak:1999rk,Butler:2000zp,Butler:2001jj,Butler:2002cw}
must be included in order to determine radiative capture and breakup cross-sections to a precision of
better than  $\sim 10\%$.

The only quantity in Eqs.~(\ref{eq:npdgsigma}) and (\ref{eq:npdgMone}) that is not determined 
by kinematics, single-nucleon properties or scattering parameters, 
is  $l_1$. In this work, 
we use LQCD to  calculate this quantity
by determining the energies of neutron-proton systems in background magnetic fields.
A magnetic field mixes the $I_z=j_z=0$ $np$ states in the $\si$ and $\siii$--$^3D_1$ channels, providing sensitivity to the EM interactions.
The deuteron and dineutron ground states are nearly degenerate at both pion masses used in the present calculation~\cite{Beane:2012vq}, 
and the two-nucleon sector exhibits an approximate spin-flavor SU(4) symmetry (as predicted  by the large-$N_c$ limit of QCD~\cite{Kaplan:1995yg}).
In this case, it can be shown \cite{polPaper} that
the energy difference between the two eigenstates
depends upon  $\tilde l_1$ as
\begin{eqnarray}
\Delta E_{^{3}S_1,^{1}S_0}({\bf B})  = 
2 \left( \kappa_1 + \gamma_0 Z_d^2 \tilde l_1 \right) {e \over M} |{\bf B}| +{\cal O}(|{\bf B}|^2)
\,,
\label{eq:Esplit}
\end{eqnarray}
where ${\bf B}$ is the background magnetic field.
It is convenient to focus on the  combination
$\overline{L}_1= \gamma_0 Z_d^2 \tilde l_1$
that characterizes the two-nucleon contributions.

Our LQCD calculations were performed on two ensembles of 
gauge-field configurations generated with a clover-improved fermion 
action~\cite{Sheikholeslami:1985ij} and  a L\"uscher-Weisz gauge 
action~\cite{Luscher:1984xn}. The first ensemble had $N_f=3$ degenerate light-quark flavors with
masses tuned to the physical strange quark mass, producing a 
pion of mass $m_\pi\sim 806~{\rm MeV}$ and used a volume of $L^3\times T=32^3\times48$.
The second ensemble had $N_f=2+1$ flavors with the same strange quark mass 
and degenerate up and down quarks with masses corresponding to a pion mass of 
$m_\pi\sim 450~{\rm MeV}$ and a volume of  $L^3\times T=32^3\times96$. Both 
ensembles had a gauge coupling of $\beta=6.1$, corresponding to a 
lattice spacing of $a \sim 0.12~{\rm fm}$. 
Background EM ($U_Q(1)$) gauge fields giving rise to 
uniform magnetic fields along the $x_3$-axis were multiplied onto each QCD 
gauge field in each ensemble (separately for each quark flavor), 
and these combined gauge fields were used to calculate up-, down-, and 
strange-quark propagators, which were then contracted to form the requisite 
nuclear correlation functions using the techniques of Ref.~\cite{Detmold:2012eu}. 
Calculations were performed on $\sim 1,000$ gauge-field configurations at the SU(3) point and $\sim 650$  
configurations at the lighter pion mass, 
each taken at  intervals of 10 hybrid Monte-Carlo trajectories. 
On each configuration, quark propagators were generated from 48 uniformly 
distributed Gaussian-smeared sources for each magnetic field. For further 
details of the production at the SU(3)-symmetric point, see Refs.~\cite{Beane:2012vq,Beane:2013br,Beane:2014ora} and 
in particular, Ref.~\cite{polPaper}. Analogous methods
were employed for the calculations using the lighter pion mass ensemble.

Background EM fields have been used extensively to 
calculate electromagnetic properties of  hadrons, such as the 
magnetic moments of the lowest-lying baryons~\cite{Bernard:1982yu,
Martinelli:1982cb,Lee:2005ds,Lee:2005dq,Detmold:2006vu,Aubin:2008qp,
Detmold:2009dx,Detmold:2010ts,Primer:2013pva} and light nuclei~\cite{Beane:2014ora},  
and the polarizabilities of mesons and baryons~\cite{Primer:2013pva,Luschevskaya:2014lga}. 
The quark fields have electric charges $Q_u=+2/3$ and $Q_{d,s} = -1/3$ 
for the up-, down- and strange-quarks, respectively, and  background 
magnetic fields are required to be quantized \cite{'tHooft:1979uj} in order that the magnetic flux is uniform throughout the lattice. The link fields, $U^{(Q)}_\mu(x)$, associated 
with the background  field are of the form
\begin{eqnarray}
U^{(Q)}_\mu(x) & = & 
e^{ i{6\pi Q_q \tilde n\over L^2} x_1 \delta_{\mu,2}} 
\times 
e^{ -i{6\pi Q_q \tilde n\over L} x_2 \delta_{\mu,1} \delta_{x_1,L-1}}
\ \ ,
\label{eq:Backfield}
\end{eqnarray}
for quark flavor $q$, where $\tilde n$ is an integer. The uniform magnetic field resulting from these links is 
$e\ {\bf B} = \ 6\pi \tilde n/L^2 \hat {\bf z}$,
where $e$ is the magnitude of the electric charge and $ \hat {\bf z}$ is a unit vector in the $x_3$-direction. 
In physical units, the background magnetic fields used with these ensembles of gauge configurations are 
$e|{\bf B}|\sim 0.05 |\tilde n|~{\rm GeV}^2$.
To optimize the re-use of light-quark propagators in the calculations,
 $U_Q(1)$ fields with $\tilde n = 0,1,-2,4$ were used.
At the SU(3) symmetric point, additional calculations were performed with $\tilde n = 3,-6,12$.

With three degenerate flavors of light quarks, and a traceless electric-charge matrix, 
there are no contributions from the magnetic field coupling to sea quarks 
at the SU(3) point at leading order in the electric charge. 
This is not the case for the $m_\pi\sim 450~{\rm MeV}$ calculations because of flavor SU(3) breaking. 
However, $\overline{L}_1$ is an isovector quantity in which 
sea quark contributions cancel (the up and down sea quarks used in this work are degenerate) so it is correctly determined by the present
calculations.

In this work, we focus on the $I_z=j_z=0$ coupled-channel neutron-proton systems. Our analysis follows 
that of Ref.~\cite{polPaper} which presents results on the $m_\pi\sim$806 MeV ensemble, and we 
direct the reader to that work for more detail regarding the interpolating operators and statistical 
analysis methods that are used.  
A matrix of correlation functions generated from source and sink operators 
associated with  $\siii$ and $\si$  $I_z=j_z=0$ interpolating operators 
\begin{figure}[!t]
	\centering
	\includegraphics[width=0.95\columnwidth]{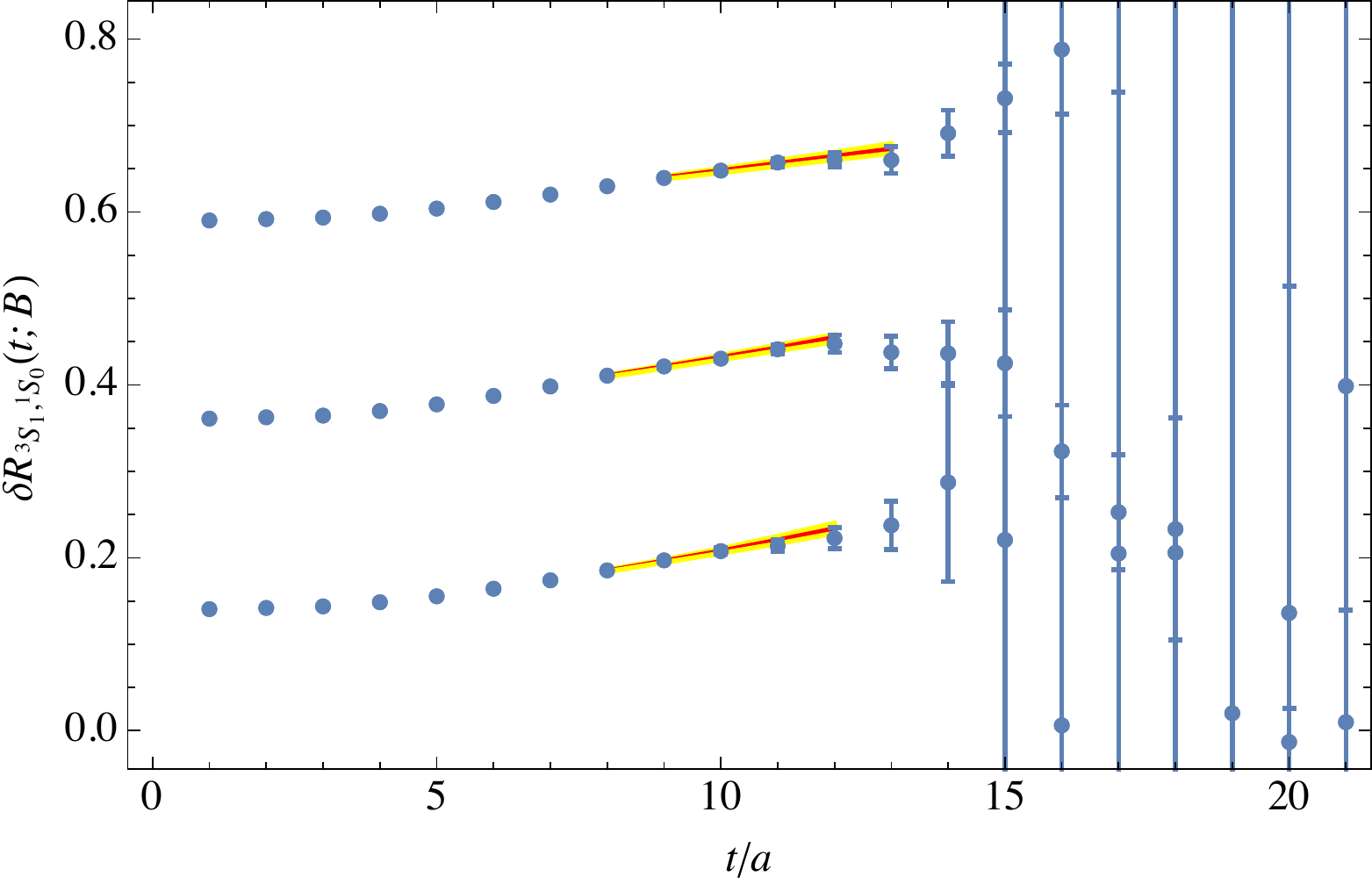} \ \
	\caption{
		The double ratios of the two principal correlators are shown for 
		$m_\pi\sim 450$ MeV for the three magnetic field strengths. The bands correspond to the 
		single-exponential fits to the correlator and the associated
		statistical uncertainty.}
	\label{fig:rat}
\end{figure}
\begin{eqnarray}
{\bf C}(t;{\bf B}) & = & 
\left(
\begin{array}{cc}
C^{\siii,\siii}(t;{\bf B}) & C^{\siii,\si}(t;{\bf B}) \\
C^{\si,\siii}(t;{\bf B}) & C^{\si,\si}(t;{\bf B}) 
\end{array}
\right)
\,,
\label{eq:cormat}
\end{eqnarray}
is diagonalized to yield ``principal correlators'', $\lambda_{\pm}(t;{\bf B})$, corresponding to the eigenstates of the 
coupled system.
In all cases, the principal correlators exhibit single-exponential behavior at times where statistical uncertainties are manageable.
  To highlight the difference arising from purely 
two-body effects, a  ratio of  ratios of the principal correlators to the appropriate 
single particle correlation functions is formed
\begin{figure}[!t]
	\centering
	\includegraphics[width=0.95\columnwidth]{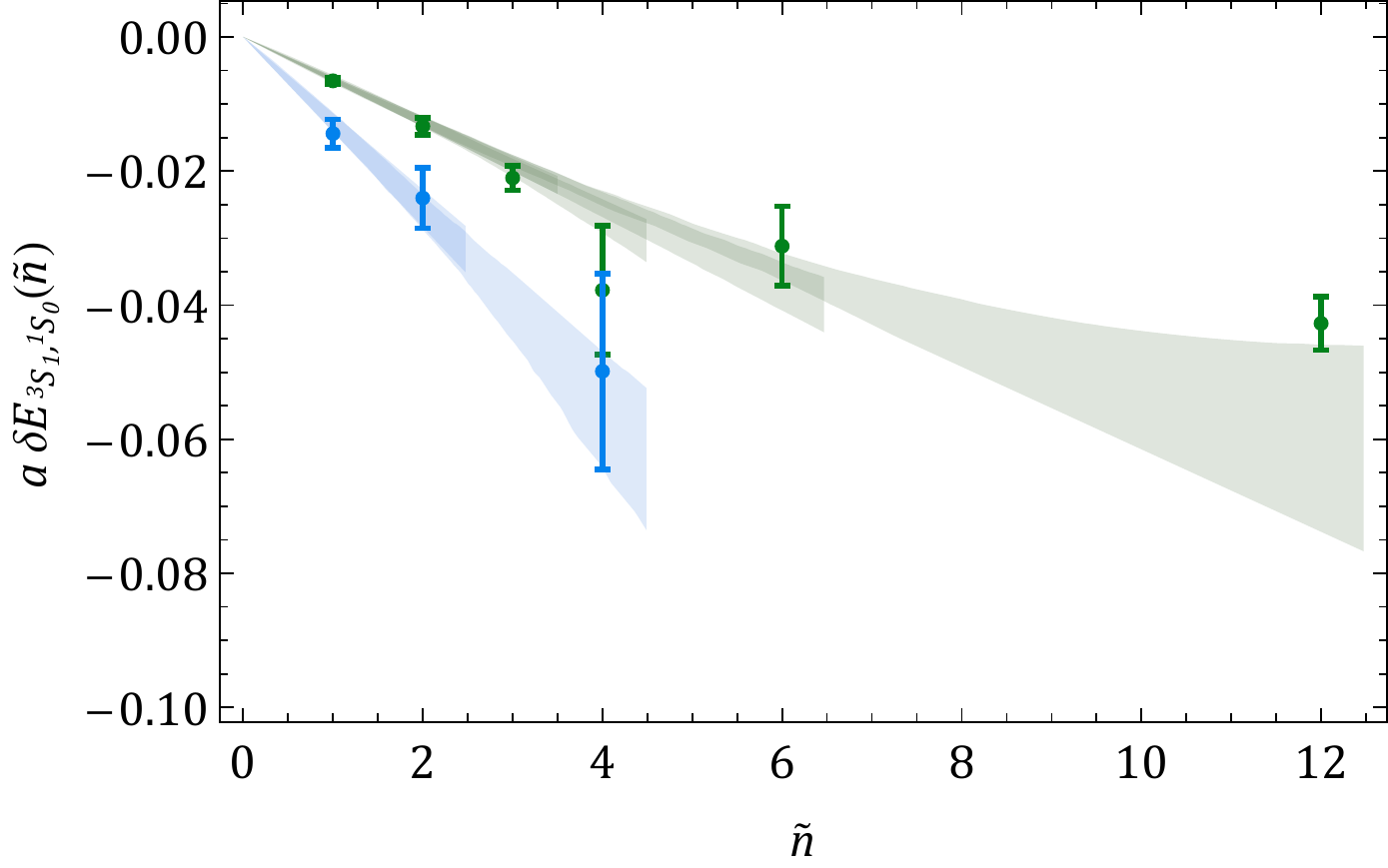} \ \
	\caption{
		LQCD calculations 
		of the energy-splittings between the two lowest-lying eigenstates, with the single-nucleon contributions removed,
		as a function of $\tilde n$, along with the associated fits. The lower (blue) set of points correspond to the 
		$m_\pi\sim 450~{\rm MeV}$ ensemble and the upper (green) points to $m_\pi\sim 806~{\rm MeV}$. The slope of the sets of points is proportional to $\overline{L}_1$ at the appropriate pion mass.
	}
	\label{fig:djz0EdeutL1}
\end{figure}
\begin{figure}[!t]
	\includegraphics[width=0.95\columnwidth]{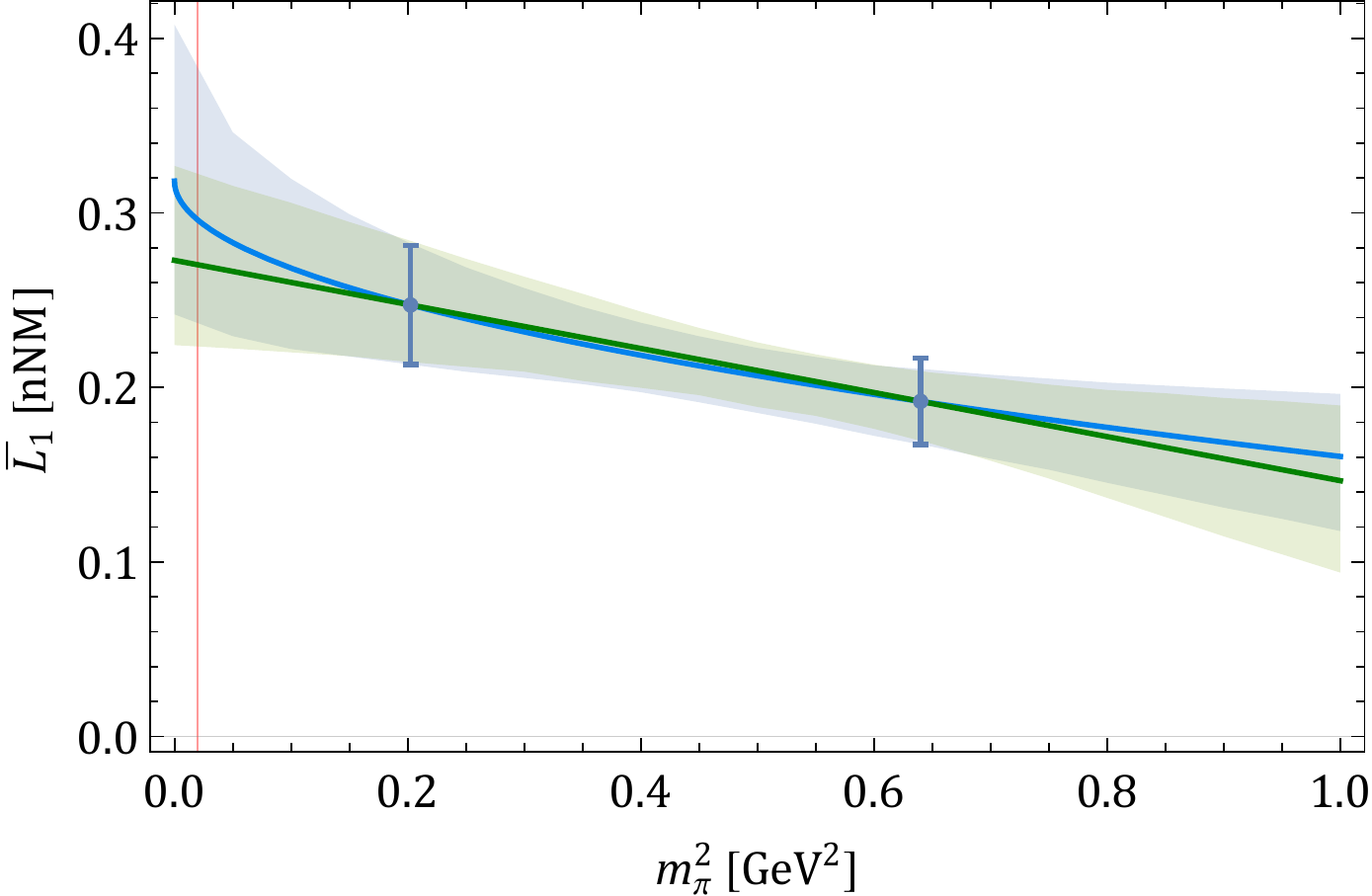} 
	\caption{
		The results of LQCD calculations of $\overline{L}_1$ (blue points).
		The blue (green) shaded regions point show the linear (quadratic) in $m_\pi$ extrapolation of $\overline{L}_1$ 
		to the physical pion mass (dashed line) in natural nuclear magnetons (nNM). The vertical (red) line indicates the physical pion mass.
	}
	\label{fig:L1extrap}
\end{figure}
\begin{eqnarray}
\delta R_{^{3}S_1,^{1}S_0}(t;{\bf B}) &=& \frac{\lambda_+(t;{\bf B})}{\lambda_-(t;{\bf B})}
\frac{C_{n,\uparrow}(t;{\bf B})C_{p,\downarrow}(t;{\bf B})}{C_{n,\downarrow}(t;{\bf B})C_{p,\uparrow}(t;{\bf B})}  
\,,
\end{eqnarray}
where $C_{p/n,\uparrow/\downarrow}(t;{\bf B})$ are the correlation functions
corresponding to the different polarizations of the proton and neutron. For large time separations, 
\begin{equation}
\delta R_{^{3}S_1,^{1}S_0}(t\to\infty;{\bf B}) \rightarrow  A\ e^{-\delta E_{^{3}S_1,^{1}S_0}({\bf B}) t},
\end{equation}
where $A$ is an overlap factor and the energy shift is
\begin{eqnarray}
\delta E_{^{3}S_1,^{1}S_0}&\equiv&\Delta E_{^{3}S_1,^{1}S_0}-[E_{p,\uparrow}-E_{p,\downarrow}]+[E_{n,\uparrow}-E_{n,\downarrow}] \nonumber \\
&\to& 2\overline{L}_1 |e {\bf B}|/M + {\cal O}({\bf B}^2)\,,
\end{eqnarray} 
omitting the ${\bf B}$ dependence for clarity.
Fig.~\ref{fig:rat} shows the above ratios  for the 
$m_\pi\sim450$~MeV ensemble for each magnetic field strength, along 
with correlated single-exponential fits to the time dependence and their statistical uncertainties.
The 
energies extracted from these fits depend on $|{\bf B}|$, with  $2\frac{e}{M}\overline{L}_1$
being the coefficient of the linear term.
Fig.~\ref{fig:djz0EdeutL1} shows the extracted energy shifts for both the $m_\pi\sim450$~MeV and $806$~MeV ensembles. The figure also shows the envelopes of a large range of polynomial  fits to their magnetic field dependence. Ref.~\cite{polPaper}
 presents the $m_\pi\sim806$~MeV correlation functions in detail, 
 and has a complete discussion of the
 fitting methods used in the analysis for both sets of pion masses.

The extracted values of $\overline{L}_1$ are shown in Fig.~\ref{fig:L1extrap} for both 
sets of quark masses. The functional dependence of $\overline{L}_1$ on the 
light-quark masses is not known. However, the deuteron and dineutron remain 
relatively near threshold over a large range of quark masses \cite{Beane:2011iw,
Yamazaki:2012hi,Beane:2012vq,Yamazaki:2015asa,Yamazaki:2015nka}, 
and the magnetic moments of the nucleons are essentially independent of the quark masses 
when expressed in units of natural nuclear magnetons \cite{Beane:2014ora}, so 
 it is plausible that $\overline{L}_1$ also varies only slowly with the pion mass.
Indeed, there is only a small difference in the value of $\overline{L}_1$
at $m_\pi\sim 806~{\rm MeV}$ and at $m_\pi\sim 450~{\rm MeV}$.
In order to connect to the physical point, we extrapolate both linearly and quadratically in the pion mass by resampling the probability distribution functions of $\overline{L}_1$ determined by the field-strength dependence fits at each pion mass.
The two forms of extrapolation yield  consistent values at the 
physical point, with the central value and uncertainties determined from  the 0.17, 0.50 and 0.83 quantiles of the combination of the two projected probability distribution functions.
After this extrapolation, the value  $\overline{L}_1^{\rm lqcd} = 0.285({\tiny \begin{array}{c}
	+63 \\ -60
	\end{array}})$~nNM 
is found at the physical pion mass, where the uncertainty incorporates statistical uncertainties, correlator fitting uncertainties, field-strength dependence fitting uncertainties, and the uncertainties in the mass extrapolation.  
This leads to a value 
$l_1^{\rm lqcd}=-4.48({\tiny \begin{array}{c}
	+16 \\ -15
	\end{array}})~{\rm fm}$.
 Future calculations with lighter quark masses will reduce both the statistical and systematic uncertainties associated with $\overline{L}_1$.

The cross section for $np\rightarrow d\gamma$ has been precisely measured in experiments 
 at an incident neutron speed of $v=2,200$~{m/s} \cite{Cox1965497}.
Using the expressions in Eqs.~(\ref{eq:npdgsigma}) and (\ref{eq:npdgMone}), 
the experimentally determined deuteron binding energy and $\si$ scattering parameters,
the experimentally determined nucleon isovector magnetic moment,
and the above extrapolated LQCD  value of $l_1^{\rm lqcd}$, leads to 
 a cross section at $v=2,200$~m/s of 
\begin{eqnarray}
\sigma^{\rm lqcd} & = & \sigmaPHYS~{\rm mb}
\ \ \ ,
\label{eq:sigPRED}
\end{eqnarray}
which is consistent 
with the experimental value of $\sigma^{\rm expt} = \sigmaEXPT~{\rm mb}$ \cite{Cox1965497} within uncertainties
(see also, Ref.~\cite{PhysRevC.74.025809}).  
As in the phenomenological determination, the two-body contributions are ${\cal O}(10\%)$. 
At the quark masses where the lattice calculations are performed, the 
cross-sections are considerably smaller than at the physical point, primarily
because the deuteron binding energy is larger.  
At $m_\pi\sim806$ MeV,  the scattering parameters, binding
energy and magnetic moments have been determined previously \cite{Beane:2012vq,Beane:2013br,Beane:2014ora} and we can predict the 
scattering cross section using only lattice QCD inputs, with a median value  $\sigma^{\rm 806\ MeV} \sim 5$ mb at $v=2,200$~m/s.\footnote{Propagation of the uncertainties in the required inputs leads to a highly non-Gaussian distribution of $\sigma^{\rm 806\ MeV}$ \cite{polPaper}.}

\vspace*{2mm}
{\it Summary:} Lattice QCD  calculations have been used to determine the short-distance 
two-nucleon interactions with the electromagnetic field (meson-exchange 
currents in the context of nuclear potential models) that make significant contributions to the low-energy cross-sections 
for $np\rightarrow d\gamma$ and $\gamma^{(*)}d\rightarrow np$. This was 
facilitated by the pionless effective field theory which provides a clean separation 
of long-distance and short-distance effects along with a concise analytic 
expression for the near-threshold cross sections. A (naive) extrapolation of the 
LQCD results to the physical pion mass is in agreement with the experimental 
determinations of the  $np\rightarrow d\gamma$ cross-section, within the 
uncertainties of the calculation and of the experiment. Calculations were 
performed at a single lattice spacing and  volume, 
introducing systematic uncertainties in $\overline{L}_1$ that are expected to be 
small in comparison to our other uncertainties, ${\cal O}(a^2\Lambda_{\rm QCD}^2, e^{-m_\pi L}, e^{-\gamma_0L})\alt 4\%$.
A more complete study, and a reduction of the uncertainties of this cross-section 
will require additional calculations at smaller lattice spacings and larger volumes, along with calculations at smaller quark masses.

The present calculation demonstrates the power of lattice QCD methods 
to address complex processes of importance to nuclear physics 
directly from the Standard Model. The methods that are used are equally applicable 
 to weak processes 
such as $pp\rightarrow d e^+\nu$, $\nu d\rightarrow ppe^+$, 
$\nu d\rightarrow \nu d$, and $\nu d\rightarrow \nu n p$, as well as  to higher-body transitions.
Background field techniques will also enable the extraction of nuclear matrix elements of other  currents relevant for searches for physics beyond the Standard Model.  Extensions of our 
studies to larger systems are currently under consideration, and 
calculations in background axial-vector fields necessary 
to address weak interaction processes are under way. As this technique has successfully recovered the short-distance contributions to $np \to d \gamma$, 
it  also seems likely that it can be generalized to the calculation of parity-violating observables in this process resulting from
 weak interactions, or from physics beyond the Standard Model (see Ref.~\cite{Schindler:2013yua} for a review).
Finally, the present work reinforces the utility of combining lattice QCD 
calculations with low-energy effective field theories describing multi-nucleon 
systems~\cite{Barnea:2013uqa}.


\ 
\begin{acknowledgments}
We are grateful to Z. Davoudi for discussions and comments. 
Calculations were performed 
using computational resources provided  by the Extreme Science and Engineering Discovery Environment (XSEDE),
which is supported by National Science Foundation grant number OCI-1053575,  
 NERSC 
(supported by U.S. Department of Energy Grant Number DE-AC02-05CH11231), 
and  by the USQCD collaboration.
This research used resources at the Oak Ridge Leadership 
  Computing Facility at the Oak Ridge National Laboratory, which is supported 
  by the Office of Science of the U.S. Department of Energy under Contract 
  No. DE-AC05-00OR22725.
Parts of the calculations made use of the {\tt chroma} software suite~\cite{Edwards:2004sx}.
SRB was partially supported by NSF
continuing grant PHY1206498 and by DOE grant DOE DE-SC0013477. EC was supported by DOE SciDAC grant DE-SC0010337-ER42045.
WD was supported by the U.S. Department of Energy Early Career Research Award DE-SC0010495. 
KO was supported by the 
U.S. Department of Energy through Grant Number DE- FG02-04ER41302 and through 
Grant Number DE-AC05-06OR23177 under which JSA operates the Thomas Jefferson 
National Accelerator Facility.  
The work of AP was supported by the contract
FIS2011-24154 from MEC (Spain) and FEDER. MJS were supported by DOE grant No.~DE-FG02-00ER41132.
BCT was supported in part by a joint City College of New York-RIKEN/Brookhaven 
Research Center fellowship, a grant from the Professional Staff Congress of the CUNY, and by the U.S. National Science Foundation, under Grant No. PHY12-05778.
\end{acknowledgments}
%

\bibliography{bib_npdg.bib}

\begin{thebibliography}{59}
\expandafter\ifx\csname natexlab\endcsname\relax\def\natexlab#1{#1}\fi
\expandafter\ifx\csname bibnamefont\endcsname\relax
  \def\bibnamefont#1{#1}\fi
\expandafter\ifx\csname bibfnamefont\endcsname\relax
  \def\bibfnamefont#1{#1}\fi
\expandafter\ifx\csname citenamefont\endcsname\relax
  \def\citenamefont#1{#1}\fi
\expandafter\ifx\csname url\endcsname\relax
  \def\url#1{\texttt{#1}}\fi
\expandafter\ifx\csname urlprefix\endcsname\relax\def\urlprefix{URL }\fi
\providecommand{\bibinfo}[2]{#2}
\providecommand{\eprint}[2][]{\url{#2}}

\bibitem[{\citenamefont{Cox et~al.}(1965)\citenamefont{Cox, Wynchank, and
  Collie}}]{Cox1965497}
\bibinfo{author}{\bibfnamefont{A.}~\bibnamefont{Cox}},
  \bibinfo{author}{\bibfnamefont{S.}~\bibnamefont{Wynchank}}, \bibnamefont{and}
  \bibinfo{author}{\bibfnamefont{C.}~\bibnamefont{Collie}},
  \bibinfo{journal}{Nuclear Physics} \textbf{\bibinfo{volume}{74}},
  \bibinfo{pages}{497 } (\bibinfo{year}{1965}).

\bibitem[{\citenamefont{Tomyo et~al.}(2003)\citenamefont{Tomyo, Nagai, Suzuki,
  Kikuchi, Shima, Kii, and Igashira}}]{Tomyo2003401}
\bibinfo{author}{\bibfnamefont{A.}~\bibnamefont{Tomyo}},
  \bibinfo{author}{\bibfnamefont{Y.}~\bibnamefont{Nagai}},
  \bibinfo{author}{\bibfnamefont{T.}~\bibnamefont{Suzuki}},
  \bibinfo{author}{\bibfnamefont{T.}~\bibnamefont{Kikuchi}},
  \bibinfo{author}{\bibfnamefont{T.}~\bibnamefont{Shima}},
  \bibinfo{author}{\bibfnamefont{T.}~\bibnamefont{Kii}}, \bibnamefont{and}
  \bibinfo{author}{\bibfnamefont{M.}~\bibnamefont{Igashira}},
  \bibinfo{journal}{Nuclear Physics A} \textbf{\bibinfo{volume}{718}},
  \bibinfo{pages}{401 } (\bibinfo{year}{2003}).

\bibitem[{\citenamefont{Nagai et~al.}(1997)\citenamefont{Nagai, Suzuki,
  Kikuchi, Shima, Kii et~al.}}]{Nagai:1997zz}
\bibinfo{author}{\bibfnamefont{Y.}~\bibnamefont{Nagai}},
  \bibinfo{author}{\bibfnamefont{T.}~\bibnamefont{Suzuki}},
  \bibinfo{author}{\bibfnamefont{T.}~\bibnamefont{Kikuchi}},
  \bibinfo{author}{\bibfnamefont{T.}~\bibnamefont{Shima}},
  \bibinfo{author}{\bibfnamefont{T.}~\bibnamefont{Kii}}, \bibnamefont{et~al.},
  \bibinfo{journal}{Phys.Rev.} \textbf{\bibinfo{volume}{C56}},
  \bibinfo{pages}{3173} (\bibinfo{year}{1997}).

\bibitem[{\citenamefont{Hara et~al.}(2003)\citenamefont{Hara, Utsunomiya, Goko,
  Akimune, Yamagata et~al.}}]{Hara:2003gw}
\bibinfo{author}{\bibfnamefont{K.~Y.} \bibnamefont{Hara}},
  \bibinfo{author}{\bibfnamefont{H.}~\bibnamefont{Utsunomiya}},
  \bibinfo{author}{\bibfnamefont{S.}~\bibnamefont{Goko}},
  \bibinfo{author}{\bibfnamefont{H.}~\bibnamefont{Akimune}},
  \bibinfo{author}{\bibfnamefont{T.}~\bibnamefont{Yamagata}},
  \bibnamefont{et~al.}, \bibinfo{journal}{Phys.Rev.}
  \textbf{\bibinfo{volume}{D68}}, \bibinfo{pages}{072001}
  (\bibinfo{year}{2003}).

\bibitem[{\citenamefont{Moreh et~al.}(1989)\citenamefont{Moreh, Kennett, and
  Prestwich}}]{Moreh:1989zz}
\bibinfo{author}{\bibfnamefont{R.}~\bibnamefont{Moreh}},
  \bibinfo{author}{\bibfnamefont{T.}~\bibnamefont{Kennett}}, \bibnamefont{and}
  \bibinfo{author}{\bibfnamefont{W.}~\bibnamefont{Prestwich}},
  \bibinfo{journal}{Phys.Rev.} \textbf{\bibinfo{volume}{C39}},
  \bibinfo{pages}{1247} (\bibinfo{year}{1989}).

\bibitem[{\citenamefont{Schreiber et~al.}(2000)\citenamefont{Schreiber, Canon,
  Crowley, Howell, Kelley et~al.}}]{Schreiber:2000tb}
\bibinfo{author}{\bibfnamefont{E.}~\bibnamefont{Schreiber}},
  \bibinfo{author}{\bibfnamefont{R.}~\bibnamefont{Canon}},
  \bibinfo{author}{\bibfnamefont{B.}~\bibnamefont{Crowley}},
  \bibinfo{author}{\bibfnamefont{C.}~\bibnamefont{Howell}},
  \bibinfo{author}{\bibfnamefont{J.}~\bibnamefont{Kelley}},
  \bibnamefont{et~al.}, \bibinfo{journal}{Phys.Rev.}
  \textbf{\bibinfo{volume}{C61}}, \bibinfo{pages}{061604}
  (\bibinfo{year}{2000}).

\bibitem[{\citenamefont{Tornow et~al.}(2003)\citenamefont{Tornow, Czakon,
  Howell, Hutcheson, Kelley et~al.}}]{Tornow:2003ze}
\bibinfo{author}{\bibfnamefont{W.}~\bibnamefont{Tornow}},
  \bibinfo{author}{\bibfnamefont{N.}~\bibnamefont{Czakon}},
  \bibinfo{author}{\bibfnamefont{C.}~\bibnamefont{Howell}},
  \bibinfo{author}{\bibfnamefont{A.}~\bibnamefont{Hutcheson}},
  \bibinfo{author}{\bibfnamefont{J.}~\bibnamefont{Kelley}},
  \bibnamefont{et~al.}, \bibinfo{journal}{Phys.Lett.}
  \textbf{\bibinfo{volume}{B574}}, \bibinfo{pages}{8} (\bibinfo{year}{2003}),
  \eprint{nucl-ex/0309009}.

\bibitem[{\citenamefont{Riska and Brown}(1972)}]{Riska:1972zz}
\bibinfo{author}{\bibfnamefont{D.}~\bibnamefont{Riska}} \bibnamefont{and}
  \bibinfo{author}{\bibfnamefont{G.}~\bibnamefont{Brown}},
  \bibinfo{journal}{Phys.Lett.} \textbf{\bibinfo{volume}{B38}},
  \bibinfo{pages}{193} (\bibinfo{year}{1972}).

\bibitem[{\citenamefont{Hockert et~al.}(1973)\citenamefont{Hockert, Riska,
  Gari, and Huffman}}]{Hockert:1974qt}
\bibinfo{author}{\bibfnamefont{J.}~\bibnamefont{Hockert}},
  \bibinfo{author}{\bibfnamefont{D.}~\bibnamefont{Riska}},
  \bibinfo{author}{\bibfnamefont{M.}~\bibnamefont{Gari}}, \bibnamefont{and}
  \bibinfo{author}{\bibfnamefont{A.}~\bibnamefont{Huffman}},
  \bibinfo{journal}{Nucl.Phys.} \textbf{\bibinfo{volume}{A217}},
  \bibinfo{pages}{14} (\bibinfo{year}{1973}).

\bibitem[{\citenamefont{Ando et~al.}(2006{\natexlab{a}})\citenamefont{Ando,
  Cyburt, Hong, and Hyun}}]{Ando:2005cz}
\bibinfo{author}{\bibfnamefont{S.}~\bibnamefont{Ando}},
  \bibinfo{author}{\bibfnamefont{R.}~\bibnamefont{Cyburt}},
  \bibinfo{author}{\bibfnamefont{S.}~\bibnamefont{Hong}}, \bibnamefont{and}
  \bibinfo{author}{\bibfnamefont{C.}~\bibnamefont{Hyun}},
  \bibinfo{journal}{Phys.Rev.} \textbf{\bibinfo{volume}{C74}},
  \bibinfo{pages}{025809} (\bibinfo{year}{2006}{\natexlab{a}}),
  \eprint{nucl-th/0511074}.

\bibitem[{\citenamefont{Kong and Ravndal}(2001)}]{Kong:2000px}
\bibinfo{author}{\bibfnamefont{X.}~\bibnamefont{Kong}} \bibnamefont{and}
  \bibinfo{author}{\bibfnamefont{F.}~\bibnamefont{Ravndal}},
  \bibinfo{journal}{Phys.Rev.} \textbf{\bibinfo{volume}{C64}},
  \bibinfo{pages}{044002} (\bibinfo{year}{2001}), \eprint{nucl-th/0004038}.

\bibitem[{\citenamefont{Butler and Chen}(2001)}]{Butler:2001jj}
\bibinfo{author}{\bibfnamefont{M.}~\bibnamefont{Butler}} \bibnamefont{and}
  \bibinfo{author}{\bibfnamefont{J.-W.} \bibnamefont{Chen}},
  \bibinfo{journal}{Phys.Lett.} \textbf{\bibinfo{volume}{B520}},
  \bibinfo{pages}{87} (\bibinfo{year}{2001}), \eprint{nucl-th/0101017}.

\bibitem[{\citenamefont{Ando et~al.}(2007)\citenamefont{Ando, Shin, Hyun, and
  Hong}}]{Ando:2007fh}
\bibinfo{author}{\bibfnamefont{S.-i.} \bibnamefont{Ando}},
  \bibinfo{author}{\bibfnamefont{J.~W.} \bibnamefont{Shin}},
  \bibinfo{author}{\bibfnamefont{C.~H.} \bibnamefont{Hyun}}, \bibnamefont{and}
  \bibinfo{author}{\bibfnamefont{S.~W.} \bibnamefont{Hong}},
  \bibinfo{journal}{Phys.Rev.} \textbf{\bibinfo{volume}{C76}},
  \bibinfo{pages}{064001} (\bibinfo{year}{2007}), \eprint{0704.2312}.

\bibitem[{\citenamefont{Adelberger et~al.}(2011)\citenamefont{Adelberger,
  Balantekin, Bemmerer, Bertulani, Chen et~al.}}]{Adelberger:2010qa}
\bibinfo{author}{\bibfnamefont{E.}~\bibnamefont{Adelberger}},
  \bibinfo{author}{\bibfnamefont{A.}~\bibnamefont{Balantekin}},
  \bibinfo{author}{\bibfnamefont{D.}~\bibnamefont{Bemmerer}},
  \bibinfo{author}{\bibfnamefont{C.}~\bibnamefont{Bertulani}},
  \bibinfo{author}{\bibfnamefont{J.-W.} \bibnamefont{Chen}},
  \bibnamefont{et~al.}, \bibinfo{journal}{Rev.Mod.Phys.}
  \textbf{\bibinfo{volume}{83}}, \bibinfo{pages}{195} (\bibinfo{year}{2011}),
  \eprint{1004.2318}.

\bibitem[{\citenamefont{Chen et~al.}(2013)\citenamefont{Chen, Liu, and
  Yu}}]{Chen:2012hm}
\bibinfo{author}{\bibfnamefont{J.-W.} \bibnamefont{Chen}},
  \bibinfo{author}{\bibfnamefont{C.-P.} \bibnamefont{Liu}}, \bibnamefont{and}
  \bibinfo{author}{\bibfnamefont{S.-H.} \bibnamefont{Yu}},
  \bibinfo{journal}{Phys.Lett.} \textbf{\bibinfo{volume}{B720}},
  \bibinfo{pages}{385} (\bibinfo{year}{2013}), \eprint{1209.2552}.

\bibitem[{\citenamefont{Marcucci et~al.}(2013)\citenamefont{Marcucci,
  Schiavilla, and Viviani}}]{Marcucci:2013tda}
\bibinfo{author}{\bibfnamefont{L.}~\bibnamefont{Marcucci}},
  \bibinfo{author}{\bibfnamefont{R.}~\bibnamefont{Schiavilla}},
  \bibnamefont{and} \bibinfo{author}{\bibfnamefont{M.}~\bibnamefont{Viviani}},
  \bibinfo{journal}{Phys.Rev.Lett.} \textbf{\bibinfo{volume}{110}},
  \bibinfo{pages}{192503} (\bibinfo{year}{2013}), \eprint{1303.3124}.

\bibitem[{\citenamefont{Rupak and Ravi}(2014)}]{Rupak:2014xza}
\bibinfo{author}{\bibfnamefont{G.}~\bibnamefont{Rupak}} \bibnamefont{and}
  \bibinfo{author}{\bibfnamefont{P.}~\bibnamefont{Ravi}},
  \bibinfo{journal}{Phys.Lett.} \textbf{\bibinfo{volume}{B741}},
  \bibinfo{pages}{301} (\bibinfo{year}{2014}), \eprint{1411.2436}.

\bibitem[{\citenamefont{Butler et~al.}(2001)\citenamefont{Butler, Chen, and
  Kong}}]{Butler:2000zp}
\bibinfo{author}{\bibfnamefont{M.}~\bibnamefont{Butler}},
  \bibinfo{author}{\bibfnamefont{J.-W.} \bibnamefont{Chen}}, \bibnamefont{and}
  \bibinfo{author}{\bibfnamefont{X.}~\bibnamefont{Kong}},
  \bibinfo{journal}{Phys.Rev.} \textbf{\bibinfo{volume}{C63}},
  \bibinfo{pages}{035501} (\bibinfo{year}{2001}), \eprint{nucl-th/0008032}.

\bibitem[{\citenamefont{Beane et~al.}(2014{\natexlab{a}})\citenamefont{Beane,
  Cohen, Detmold, Lin, and Savage}}]{Beane:2013kca}
\bibinfo{author}{\bibfnamefont{S.}~\bibnamefont{Beane}},
  \bibinfo{author}{\bibfnamefont{S.}~\bibnamefont{Cohen}},
  \bibinfo{author}{\bibfnamefont{W.}~\bibnamefont{Detmold}},
  \bibinfo{author}{\bibfnamefont{H.~W.} \bibnamefont{Lin}}, \bibnamefont{and}
  \bibinfo{author}{\bibfnamefont{M.}~\bibnamefont{Savage}},
  \bibinfo{journal}{Phys.Rev.} \textbf{\bibinfo{volume}{D89}},
  \bibinfo{pages}{074505} (\bibinfo{year}{2014}{\natexlab{a}}),
  \eprint{1306.6939}.

\bibitem[{\citenamefont{Bethe and Longmire}(1950)}]{Bethe:1950jm}
\bibinfo{author}{\bibfnamefont{H.}~\bibnamefont{Bethe}} \bibnamefont{and}
  \bibinfo{author}{\bibfnamefont{C.}~\bibnamefont{Longmire}},
  \bibinfo{journal}{Phys.Rev.} \textbf{\bibinfo{volume}{77}},
  \bibinfo{pages}{647} (\bibinfo{year}{1950}).

\bibitem[{\citenamefont{Noyes}(1965)}]{Noyes:1964fx}
\bibinfo{author}{\bibfnamefont{H.~P.} \bibnamefont{Noyes}},
  \bibinfo{journal}{Nucl.Phys.} \textbf{\bibinfo{volume}{74}},
  \bibinfo{pages}{508} (\bibinfo{year}{1965}).

\bibitem[{\citenamefont{Rupak}(2000)}]{Rupak:1999rk}
\bibinfo{author}{\bibfnamefont{G.}~\bibnamefont{Rupak}},
  \bibinfo{journal}{Nucl.Phys.} \textbf{\bibinfo{volume}{A678}},
  \bibinfo{pages}{405} (\bibinfo{year}{2000}), \eprint{nucl-th/9911018}.

\bibitem[{\citenamefont{Kaplan et~al.}(1999)\citenamefont{Kaplan, Savage, and
  Wise}}]{Kaplan:1998sz}
\bibinfo{author}{\bibfnamefont{D.~B.} \bibnamefont{Kaplan}},
  \bibinfo{author}{\bibfnamefont{M.~J.} \bibnamefont{Savage}},
  \bibnamefont{and} \bibinfo{author}{\bibfnamefont{M.~B.} \bibnamefont{Wise}},
  \bibinfo{journal}{Phys.Rev.} \textbf{\bibinfo{volume}{C59}},
  \bibinfo{pages}{617} (\bibinfo{year}{1999}), \eprint{nucl-th/9804032}.

\bibitem[{\citenamefont{Kaplan et~al.}(1998)\citenamefont{Kaplan, Savage, and
  Wise}}]{Kaplan:1998we}
\bibinfo{author}{\bibfnamefont{D.~B.} \bibnamefont{Kaplan}},
  \bibinfo{author}{\bibfnamefont{M.~J.} \bibnamefont{Savage}},
  \bibnamefont{and} \bibinfo{author}{\bibfnamefont{M.~B.} \bibnamefont{Wise}},
  \bibinfo{journal}{Nucl.Phys.} \textbf{\bibinfo{volume}{B534}},
  \bibinfo{pages}{329} (\bibinfo{year}{1998}), \eprint{nucl-th/9802075}.

\bibitem[{\citenamefont{van Kolck}(1999)}]{vanKolck:1998bw}
\bibinfo{author}{\bibfnamefont{U.}~\bibnamefont{van Kolck}},
  \bibinfo{journal}{Nucl.Phys.} \textbf{\bibinfo{volume}{A645}},
  \bibinfo{pages}{273} (\bibinfo{year}{1999}), \eprint{nucl-th/9808007}.

\bibitem[{\citenamefont{Kaplan}(1997)}]{Kaplan:1996nv}
\bibinfo{author}{\bibfnamefont{D.~B.} \bibnamefont{Kaplan}},
  \bibinfo{journal}{Nucl.Phys.} \textbf{\bibinfo{volume}{B494}},
  \bibinfo{pages}{471} (\bibinfo{year}{1997}), \eprint{nucl-th/9610052}.

\bibitem[{\citenamefont{Beane and Savage}(2001)}]{Beane:2000fi}
\bibinfo{author}{\bibfnamefont{S.~R.} \bibnamefont{Beane}} \bibnamefont{and}
  \bibinfo{author}{\bibfnamefont{M.~J.} \bibnamefont{Savage}},
  \bibinfo{journal}{Nucl.Phys.} \textbf{\bibinfo{volume}{A694}},
  \bibinfo{pages}{511} (\bibinfo{year}{2001}), \eprint{nucl-th/0011067}.

\bibitem[{\citenamefont{Detmold and Savage}(2004)}]{Detmold:2004qn}
\bibinfo{author}{\bibfnamefont{W.}~\bibnamefont{Detmold}} \bibnamefont{and}
  \bibinfo{author}{\bibfnamefont{M.~J.} \bibnamefont{Savage}},
  \bibinfo{journal}{Nucl.Phys.} \textbf{\bibinfo{volume}{A743}},
  \bibinfo{pages}{170} (\bibinfo{year}{2004}), \eprint{hep-lat/0403005}.

\bibitem[{\citenamefont{Chen et~al.}(1999{\natexlab{a}})\citenamefont{Chen,
  Rupak, and Savage}}]{Chen:1999tn}
\bibinfo{author}{\bibfnamefont{J.-W.} \bibnamefont{Chen}},
  \bibinfo{author}{\bibfnamefont{G.}~\bibnamefont{Rupak}}, \bibnamefont{and}
  \bibinfo{author}{\bibfnamefont{M.~J.} \bibnamefont{Savage}},
  \bibinfo{journal}{Nucl.Phys.} \textbf{\bibinfo{volume}{A653}},
  \bibinfo{pages}{386} (\bibinfo{year}{1999}{\natexlab{a}}),
  \eprint{nucl-th/9902056}.

\bibitem[{\citenamefont{Chen et~al.}(1999{\natexlab{b}})\citenamefont{Chen,
  Rupak, and Savage}}]{Chen:1999vd}
\bibinfo{author}{\bibfnamefont{J.-W.} \bibnamefont{Chen}},
  \bibinfo{author}{\bibfnamefont{G.}~\bibnamefont{Rupak}}, \bibnamefont{and}
  \bibinfo{author}{\bibfnamefont{M.~J.} \bibnamefont{Savage}},
  \bibinfo{journal}{Phys.Lett.} \textbf{\bibinfo{volume}{B464}},
  \bibinfo{pages}{1} (\bibinfo{year}{1999}{\natexlab{b}}),
  \eprint{nucl-th/9905002}.

\bibitem[{\citenamefont{Butler and Chen}(2000)}]{Butler:1999sv}
\bibinfo{author}{\bibfnamefont{M.}~\bibnamefont{Butler}} \bibnamefont{and}
  \bibinfo{author}{\bibfnamefont{J.-W.} \bibnamefont{Chen}},
  \bibinfo{journal}{Nucl.Phys.} \textbf{\bibinfo{volume}{A675}},
  \bibinfo{pages}{575} (\bibinfo{year}{2000}), \eprint{nucl-th/9905059}.

\bibitem[{\citenamefont{Butler et~al.}(2002)\citenamefont{Butler, Chen, and
  Vogel}}]{Butler:2002cw}
\bibinfo{author}{\bibfnamefont{M.}~\bibnamefont{Butler}},
  \bibinfo{author}{\bibfnamefont{J.-W.} \bibnamefont{Chen}}, \bibnamefont{and}
  \bibinfo{author}{\bibfnamefont{P.}~\bibnamefont{Vogel}},
  \bibinfo{journal}{Phys.Lett.} \textbf{\bibinfo{volume}{B549}},
  \bibinfo{pages}{26} (\bibinfo{year}{2002}), \eprint{nucl-th/0206026}.

\bibitem[{\citenamefont{Beane et~al.}(2013{\natexlab{a}})}]{Beane:2012vq}
\bibinfo{author}{\bibfnamefont{S.}~\bibnamefont{Beane}} \bibnamefont{et~al.}
  (\bibinfo{collaboration}{NPLQCD}), \bibinfo{journal}{Phys.Rev.}
  \textbf{\bibinfo{volume}{D87}}, \bibinfo{pages}{034506}
  (\bibinfo{year}{2013}{\natexlab{a}}), \eprint{1206.5219}.

\bibitem[{\citenamefont{Kaplan and Savage}(1996)}]{Kaplan:1995yg}
\bibinfo{author}{\bibfnamefont{D.~B.} \bibnamefont{Kaplan}} \bibnamefont{and}
  \bibinfo{author}{\bibfnamefont{M.~J.} \bibnamefont{Savage}},
  \bibinfo{journal}{Phys.Lett.} \textbf{\bibinfo{volume}{B365}},
  \bibinfo{pages}{244} (\bibinfo{year}{1996}), \eprint{hep-ph/9509371}.

\bibitem[{\citenamefont{Beane et~al.}()\citenamefont{Beane, Chang, Detmold,
  Orginos, Parre{\~n}o, Savage, and Tiburzi}}]{polPaper}
\bibinfo{author}{\bibfnamefont{S.}~\bibnamefont{Beane}},
  \bibinfo{author}{\bibfnamefont{E.}~\bibnamefont{Chang}},
  \bibinfo{author}{\bibfnamefont{W.}~\bibnamefont{Detmold}},
  \bibinfo{author}{\bibfnamefont{K.}~\bibnamefont{Orginos}},
  \bibinfo{author}{\bibfnamefont{A.}~\bibnamefont{Parre{\~n}o}},
  \bibinfo{author}{\bibfnamefont{M.~J.} \bibnamefont{Savage}},
  \bibnamefont{and} \bibinfo{author}{\bibfnamefont{B.~C.}
  \bibnamefont{Tiburzi}}, \emph{\bibinfo{title}{{Magnetic properties of light
  nuclei from lattice quantum chromodynamics}}}, \bibinfo{note}{to appear}.

\bibitem[{\citenamefont{Sheikholeslami and
  Wohlert}(1985)}]{Sheikholeslami:1985ij}
\bibinfo{author}{\bibfnamefont{B.}~\bibnamefont{Sheikholeslami}}
  \bibnamefont{and} \bibinfo{author}{\bibfnamefont{R.}~\bibnamefont{Wohlert}},
  \bibinfo{journal}{Nucl.Phys.} \textbf{\bibinfo{volume}{B259}},
  \bibinfo{pages}{572} (\bibinfo{year}{1985}).

\bibitem[{\citenamefont{L{\"u}scher and Weisz}(1985)}]{Luscher:1984xn}
\bibinfo{author}{\bibfnamefont{M.}~\bibnamefont{L{\"u}scher}} \bibnamefont{and}
  \bibinfo{author}{\bibfnamefont{P.}~\bibnamefont{Weisz}},
  \bibinfo{journal}{Commun.Math.Phys.} \textbf{\bibinfo{volume}{97}},
  \bibinfo{pages}{59} (\bibinfo{year}{1985}).

\bibitem[{\citenamefont{Detmold and Orginos}(2013)}]{Detmold:2012eu}
\bibinfo{author}{\bibfnamefont{W.}~\bibnamefont{Detmold}} \bibnamefont{and}
  \bibinfo{author}{\bibfnamefont{K.}~\bibnamefont{Orginos}},
  \bibinfo{journal}{Phys.Rev.} \textbf{\bibinfo{volume}{D87}},
  \bibinfo{pages}{114512} (\bibinfo{year}{2013}), \eprint{1207.1452}.

\bibitem[{\citenamefont{Beane et~al.}(2013{\natexlab{b}})}]{Beane:2013br}
\bibinfo{author}{\bibfnamefont{S.}~\bibnamefont{Beane}} \bibnamefont{et~al.}
  (\bibinfo{collaboration}{NPLQCD}), \bibinfo{journal}{Phys.Rev.}
  \textbf{\bibinfo{volume}{C88}}, \bibinfo{pages}{024003}
  (\bibinfo{year}{2013}{\natexlab{b}}), \eprint{1301.5790}.

\bibitem[{\citenamefont{Beane et~al.}(2014{\natexlab{b}})\citenamefont{Beane,
  Chang, Cohen, Detmold, Lin et~al.}}]{Beane:2014ora}
\bibinfo{author}{\bibfnamefont{S.}~\bibnamefont{Beane}},
  \bibinfo{author}{\bibfnamefont{E.}~\bibnamefont{Chang}},
  \bibinfo{author}{\bibfnamefont{S.}~\bibnamefont{Cohen}},
  \bibinfo{author}{\bibfnamefont{W.}~\bibnamefont{Detmold}},
  \bibinfo{author}{\bibfnamefont{H.}~\bibnamefont{Lin}}, \bibnamefont{et~al.},
  \bibinfo{journal}{Phys.Rev.Lett.} \textbf{\bibinfo{volume}{113}},
  \bibinfo{pages}{252001} (\bibinfo{year}{2014}{\natexlab{b}}),
  \eprint{1409.3556}.

\bibitem[{\citenamefont{Bernard et~al.}(1982)\citenamefont{Bernard, Draper,
  Olynyk, and Rushton}}]{Bernard:1982yu}
\bibinfo{author}{\bibfnamefont{C.~W.} \bibnamefont{Bernard}},
  \bibinfo{author}{\bibfnamefont{T.}~\bibnamefont{Draper}},
  \bibinfo{author}{\bibfnamefont{K.}~\bibnamefont{Olynyk}}, \bibnamefont{and}
  \bibinfo{author}{\bibfnamefont{M.}~\bibnamefont{Rushton}},
  \bibinfo{journal}{Phys.Rev.Lett.} \textbf{\bibinfo{volume}{49}},
  \bibinfo{pages}{1076} (\bibinfo{year}{1982}).

\bibitem[{\citenamefont{Martinelli et~al.}(1982)\citenamefont{Martinelli,
  Parisi, Petronzio, and Rapuano}}]{Martinelli:1982cb}
\bibinfo{author}{\bibfnamefont{G.}~\bibnamefont{Martinelli}},
  \bibinfo{author}{\bibfnamefont{G.}~\bibnamefont{Parisi}},
  \bibinfo{author}{\bibfnamefont{R.}~\bibnamefont{Petronzio}},
  \bibnamefont{and} \bibinfo{author}{\bibfnamefont{F.}~\bibnamefont{Rapuano}},
  \bibinfo{journal}{Phys.Lett.} \textbf{\bibinfo{volume}{B116}},
  \bibinfo{pages}{434} (\bibinfo{year}{1982}).

\bibitem[{\citenamefont{Lee et~al.}(2005)\citenamefont{Lee, Kelly, Zhou, and
  Wilcox}}]{Lee:2005ds}
\bibinfo{author}{\bibfnamefont{F.}~\bibnamefont{Lee}},
  \bibinfo{author}{\bibfnamefont{R.}~\bibnamefont{Kelly}},
  \bibinfo{author}{\bibfnamefont{L.}~\bibnamefont{Zhou}}, \bibnamefont{and}
  \bibinfo{author}{\bibfnamefont{W.}~\bibnamefont{Wilcox}},
  \bibinfo{journal}{Phys.Lett.} \textbf{\bibinfo{volume}{B627}},
  \bibinfo{pages}{71} (\bibinfo{year}{2005}), \eprint{hep-lat/0509067}.

\bibitem[{\citenamefont{Lee et~al.}(2006)\citenamefont{Lee, Zhou, Wilcox, and
  Christensen}}]{Lee:2005dq}
\bibinfo{author}{\bibfnamefont{F.~X.} \bibnamefont{Lee}},
  \bibinfo{author}{\bibfnamefont{L.}~\bibnamefont{Zhou}},
  \bibinfo{author}{\bibfnamefont{W.}~\bibnamefont{Wilcox}}, \bibnamefont{and}
  \bibinfo{author}{\bibfnamefont{J.~C.} \bibnamefont{Christensen}},
  \bibinfo{journal}{Phys.Rev.} \textbf{\bibinfo{volume}{D73}},
  \bibinfo{pages}{034503} (\bibinfo{year}{2006}), \eprint{hep-lat/0509065}.

\bibitem[{\citenamefont{Detmold et~al.}(2006)\citenamefont{Detmold, Tiburzi,
  and Walker-Loud}}]{Detmold:2006vu}
\bibinfo{author}{\bibfnamefont{W.}~\bibnamefont{Detmold}},
  \bibinfo{author}{\bibfnamefont{B.}~\bibnamefont{Tiburzi}}, \bibnamefont{and}
  \bibinfo{author}{\bibfnamefont{A.}~\bibnamefont{Walker-Loud}},
  \bibinfo{journal}{Phys.Rev.} \textbf{\bibinfo{volume}{D73}},
  \bibinfo{pages}{114505} (\bibinfo{year}{2006}), \eprint{hep-lat/0603026}.

\bibitem[{\citenamefont{Aubin et~al.}(2009)\citenamefont{Aubin, Orginos,
  Pascalutsa, and Vanderhaeghen}}]{Aubin:2008qp}
\bibinfo{author}{\bibfnamefont{C.}~\bibnamefont{Aubin}},
  \bibinfo{author}{\bibfnamefont{K.}~\bibnamefont{Orginos}},
  \bibinfo{author}{\bibfnamefont{V.}~\bibnamefont{Pascalutsa}},
  \bibnamefont{and}
  \bibinfo{author}{\bibfnamefont{M.}~\bibnamefont{Vanderhaeghen}},
  \bibinfo{journal}{Phys.Rev.} \textbf{\bibinfo{volume}{D79}},
  \bibinfo{pages}{051502} (\bibinfo{year}{2009}), \eprint{0811.2440}.

\bibitem[{\citenamefont{Detmold et~al.}(2009)\citenamefont{Detmold, Tiburzi,
  and Walker-Loud}}]{Detmold:2009dx}
\bibinfo{author}{\bibfnamefont{W.}~\bibnamefont{Detmold}},
  \bibinfo{author}{\bibfnamefont{B.~C.} \bibnamefont{Tiburzi}},
  \bibnamefont{and}
  \bibinfo{author}{\bibfnamefont{A.}~\bibnamefont{Walker-Loud}},
  \bibinfo{journal}{Phys.Rev.} \textbf{\bibinfo{volume}{D79}},
  \bibinfo{pages}{094505} (\bibinfo{year}{2009}), \eprint{0904.1586}.

\bibitem[{\citenamefont{Detmold et~al.}(2010)\citenamefont{Detmold, Tiburzi,
  and Walker-Loud}}]{Detmold:2010ts}
\bibinfo{author}{\bibfnamefont{W.}~\bibnamefont{Detmold}},
  \bibinfo{author}{\bibfnamefont{B.}~\bibnamefont{Tiburzi}}, \bibnamefont{and}
  \bibinfo{author}{\bibfnamefont{A.}~\bibnamefont{Walker-Loud}},
  \bibinfo{journal}{Phys.Rev.} \textbf{\bibinfo{volume}{D81}},
  \bibinfo{pages}{054502} (\bibinfo{year}{2010}), \eprint{1001.1131}.

\bibitem[{\citenamefont{Primer et~al.}(2014)\citenamefont{Primer, Kamleh,
  Leinweber, and Burkardt}}]{Primer:2013pva}
\bibinfo{author}{\bibfnamefont{T.}~\bibnamefont{Primer}},
  \bibinfo{author}{\bibfnamefont{W.}~\bibnamefont{Kamleh}},
  \bibinfo{author}{\bibfnamefont{D.}~\bibnamefont{Leinweber}},
  \bibnamefont{and} \bibinfo{author}{\bibfnamefont{M.}~\bibnamefont{Burkardt}},
  \bibinfo{journal}{Phys.Rev.} \textbf{\bibinfo{volume}{D89}},
  \bibinfo{pages}{034508} (\bibinfo{year}{2014}), \eprint{1307.1509}.

\bibitem[{\citenamefont{Luschevskaya et~al.}(2014)\citenamefont{Luschevskaya,
  Teryaev, and Kochetkov}}]{Luschevskaya:2014lga}
\bibinfo{author}{\bibfnamefont{E.}~\bibnamefont{Luschevskaya}},
  \bibinfo{author}{\bibfnamefont{O.}~\bibnamefont{Teryaev}}, \bibnamefont{and}
  \bibinfo{author}{\bibfnamefont{O.}~\bibnamefont{Kochetkov}}
  (\bibinfo{year}{2014}), \eprint{1411.4284}.

\bibitem[{\citenamefont{'t~Hooft}(1979)}]{'tHooft:1979uj}
\bibinfo{author}{\bibfnamefont{G.}~\bibnamefont{'t~Hooft}},
  \bibinfo{journal}{Nucl.Phys.} \textbf{\bibinfo{volume}{B153}},
  \bibinfo{pages}{141} (\bibinfo{year}{1979}).

\bibitem[{\citenamefont{Beane et~al.}(2012)}]{Beane:2011iw}
\bibinfo{author}{\bibfnamefont{S.}~\bibnamefont{Beane}} \bibnamefont{et~al.}
  (\bibinfo{collaboration}{NPLQCD}), \bibinfo{journal}{Phys.Rev.}
  \textbf{\bibinfo{volume}{D85}}, \bibinfo{pages}{054511}
  (\bibinfo{year}{2012}), \eprint{1109.2889}.

\bibitem[{\citenamefont{Yamazaki et~al.}(2012)\citenamefont{Yamazaki, Ishikawa,
  Kuramashi, and Ukawa}}]{Yamazaki:2012hi}
\bibinfo{author}{\bibfnamefont{T.}~\bibnamefont{Yamazaki}},
  \bibinfo{author}{\bibfnamefont{K.-i.} \bibnamefont{Ishikawa}},
  \bibinfo{author}{\bibfnamefont{Y.}~\bibnamefont{Kuramashi}},
  \bibnamefont{and} \bibinfo{author}{\bibfnamefont{A.}~\bibnamefont{Ukawa}},
  \bibinfo{journal}{Phys.Rev.} \textbf{\bibinfo{volume}{D86}},
  \bibinfo{pages}{074514} (\bibinfo{year}{2012}), \eprint{1207.4277}.

\bibitem[{\citenamefont{Yamazaki et~al.}(2015)\citenamefont{Yamazaki, Ishikawa,
  Kuramashi, and Ukawa}}]{Yamazaki:2015asa}
\bibinfo{author}{\bibfnamefont{T.}~\bibnamefont{Yamazaki}},
  \bibinfo{author}{\bibfnamefont{K.-i.} \bibnamefont{Ishikawa}},
  \bibinfo{author}{\bibfnamefont{Y.}~\bibnamefont{Kuramashi}},
  \bibnamefont{and} \bibinfo{author}{\bibfnamefont{A.}~\bibnamefont{Ukawa}}
  (\bibinfo{year}{2015}), \eprint{1502.04182}.

\bibitem[{\citenamefont{Yamazaki}(2015)}]{Yamazaki:2015nka}
\bibinfo{author}{\bibfnamefont{T.}~\bibnamefont{Yamazaki}}
  (\bibinfo{year}{2015}), \eprint{1503.08671}.

\bibitem[{\citenamefont{Ando et~al.}(2006{\natexlab{b}})\citenamefont{Ando,
  Cyburt, Hong, and Hyun}}]{PhysRevC.74.025809}
\bibinfo{author}{\bibfnamefont{S.}~\bibnamefont{Ando}},
  \bibinfo{author}{\bibfnamefont{R.~H.} \bibnamefont{Cyburt}},
  \bibinfo{author}{\bibfnamefont{S.~W.} \bibnamefont{Hong}}, \bibnamefont{and}
  \bibinfo{author}{\bibfnamefont{C.~H.} \bibnamefont{Hyun}},
  \bibinfo{journal}{Phys. Rev. C} \textbf{\bibinfo{volume}{74}},
  \bibinfo{pages}{025809} (\bibinfo{year}{2006}{\natexlab{b}}).

\bibitem[{\citenamefont{Schindler and Springer}(2013)}]{Schindler:2013yua}
\bibinfo{author}{\bibfnamefont{M.}~\bibnamefont{Schindler}} \bibnamefont{and}
  \bibinfo{author}{\bibfnamefont{R.}~\bibnamefont{Springer}},
  \bibinfo{journal}{Prog.Part.Nucl.Phys.} \textbf{\bibinfo{volume}{72}},
  \bibinfo{pages}{1} (\bibinfo{year}{2013}), \eprint{1305.4190}.

\bibitem[{\citenamefont{Barnea et~al.}(2015)\citenamefont{Barnea, Contessi,
  Gazit, Pederiva, and van Kolck}}]{Barnea:2013uqa}
\bibinfo{author}{\bibfnamefont{N.}~\bibnamefont{Barnea}},
  \bibinfo{author}{\bibfnamefont{L.}~\bibnamefont{Contessi}},
  \bibinfo{author}{\bibfnamefont{D.}~\bibnamefont{Gazit}},
  \bibinfo{author}{\bibfnamefont{F.}~\bibnamefont{Pederiva}}, \bibnamefont{and}
  \bibinfo{author}{\bibfnamefont{U.}~\bibnamefont{van Kolck}},
  \bibinfo{journal}{Phys.Rev.Lett.} \textbf{\bibinfo{volume}{114}},
  \bibinfo{pages}{052501} (\bibinfo{year}{2015}), \eprint{1311.4966}.

\bibitem[{\citenamefont{Edwards and Joo}(2005)}]{Edwards:2004sx}
\bibinfo{author}{\bibfnamefont{R.~G.} \bibnamefont{Edwards}} \bibnamefont{and}
  \bibinfo{author}{\bibfnamefont{B.}~\bibnamefont{Joo}}
  (\bibinfo{collaboration}{SciDAC Collaboration, LHPC Collaboration, UKQCD
  Collaboration}), \bibinfo{journal}{Nucl.Phys.Proc.Suppl.}
  \textbf{\bibinfo{volume}{140}}, \bibinfo{pages}{832} (\bibinfo{year}{2005}),
  \eprint{hep-lat/0409003}.

\end{thebibliography}
\end{document}